\documentclass[runningheads]{llncs}
\usepackage[T1]{fontenc}
\usepackage{tabularx}
\usepackage{graphicx}
\usepackage{float}
\usepackage{hyperref}
\DeclareUnicodeCharacter{2212}{-}
\begin{document}
\title{Adversarial Debiasing of Machine Learning Models for Enhanced Network Security against DDoS Attacks}
\titlerunning{Adversarial Debiasing  for Enhanced Network Security}
\authorrunning{A. Sukumar et al.}
%
\author{Aadith Sukumar\inst{1}\and
Isha Singh\inst{1} \and
Devershika Mohane\inst{1} \and
Ankit Mukherjee\inst{1} \and
Ankush Dutta\inst{1} \and
Rahee Walambe*\inst{1,2} \and
Ketan Kotecha\inst{1,2,3}}

\institute{Symbiosis Institute of Technology Pune, India \and
Symbiosis Center for Applied Artificial Intelligence, Pune, India \and
Peoples’ Friendship University of Russia Moscow, Russian Federation}
\maketitle              
\begin{abstract}
Distributed Denial of Service (DDoS) attacks are a growing threat to network infrastructure, and new techniques, including the use of generative AI, make them harder to detect. Traditional detection systems, such as rule-based firewalls, often fail to identify these evolving attack patterns. In this study, we propose a new method for detecting DDoS attacks by combining synthetic data generation using Generative Adversarial Networks (GANs) with a Random Forest classifier. The GAN-generated data showed 80.3\% cosine similarity to real traffic, which helped the model learn underlying traffic patterns more effectively. To address imbalances in the data, especially in packet-related features, we applied adversarial debiasing. This reduced the model’s sensitivity to skewed distributions in variables such as forward and backward packet counts and total byte lengths. Our results show that models trained on a mix of synthetic and real data achieved significantly better performance: 99.98\% accuracy on benchmark data and a 22.60\% improvement when tested on previously unseen synthetic traffic. This suggests that the method can generalize well across different traffic scenarios and adapt quickly to new types of attacks. The proposed approach not only improves DDoS detection but also provides a scalable foundation for security models that account for bias and benefit from data augmentation. Our findings show that combining GANs with adversarial debiasing can lead to more robust and effective DDoS mitigation, supporting the further development of machine learning–based cyber security.

\end{abstract}
\section{Introduction}
In the current digital era, Distributed Denial of Service (DDoS) attacks have emerged as one of the most formidable threats to network security. DDoS attacks target to interfere with the availability of network services by flooding systems with malicious traffic. These attacks are typically launched from a large network of compromised devices, known as a botnet, which simultaneously sends massive requests to the target. DDoS can also be used as a justification for many malicious activities, such as turning off security measures and breaking through the target's boundary [1]. This results in restricting access to its resources until the service is severely compromised.

DDoS attacks are broadly classified into three main types: volumetric attacks, protocol attacks, and application layer attacks. These attacks include the User Datagram Protocol flooding, where a large volume of data is sent into a network to fill up the available bandwidth, hence denying other traffic. Vulnerabilities in network protocols are exploited in protocol attacks e.g., SYN Flooding. The application layer attacks, such as HTTP flooding, are more sophisticated. They mimic legitimate traffic at the application level [2]. These attacks differ from legitimate attacks, making it hard to detect. Each of these attacks targets different aspects of the network, requiring unique detection methods. 

The study and solutions for these attacks are of interest to the research community for last few years. A number of studies and methods have been proposed for various environments. Classification techniques, especially those belonging to the category of machine learning algorithms, are widely used for the detection of DDoS attacks since they sort network traffic according to previously defined classes. 

Random Forest is one of the effective and robust ensemble-learning methods for distinguishing between legitimate and malicious traffic [2]. Various approaches [[2-4]] employing multiple AI models for attack detection are reported in the literature. In [3], a hybrid, incremental-learning framework for DDoS detection using supervised classifiers is discussed. The proposed system splits processing between client and proxy and applies multiple learners (e.g. Naïve Bayes, decision trees, MLP, k-NN) to detect attacks, with Random Forest emerging as the strongest performer. This confirms that ensemble methods like RF are effective in DDoS classification. However, this hybrid approach focuses on accuracy and processing speed in streaming data rather than on data imbalance or bias. The paper does not employ any generative data augmentation or adversarial training; it relies solely on the original (often imbalanced) samples. The data imbalance problem is explicitly tackled in [5] for IoT-driven DDoS detection. They identify two extreme scenarios – low-rate versus high-rate DDoS attacks – and show that classifiers suffer when cross-evaluated across such varied datasets. To counter this, they experiment with preprocessing techniques and conclude that oversampling should be used for low-rate (rare) attacks and undersampling for high-rate attacks, even suggesting optimized SMOTE oversamplers per model. In essence, they recognize that two distinct scenarios of data imbalance exist and that addressing both is crucial. Agrawal et al. [2] conduct a broad benchmarking of ML classifiers for DDoS detection. Using four diverse public datasets (e.g. CICDDoS19, backscatter TCP/ICMP, SDN logs) they compare algorithms like CatBoost, logistic regression, gradient boosting and Naïve Bayes on accuracy, precision, recall and F1 score. However, this work does not address data bias or imbalance. An adversarial deep-learning system for DDoS detection is proposed in [6,7)]. Novaes et al [6], employ a Generative Adversarial Network (GAN) framework to not only detect DDoS flows but also to simulate adversarial DDoS traffic for training. Through adversarial training, their model becomes “less sensitive to adversarial attacks” on the network, improving robustness against evasion. In recent times, generative AI techniques have also been proposed [8, 9]. Genetic algorithms and a fuzzy logic-based approach are reported in [10]. For cloud-based environments, various methods and techniques for intrusion and attack detection are proposed in literature [11-15]. For the SDN environment, multiple ML and Deep learning based methods are reported [16-22]. The data bias issue for deploying machine learning models for such tasks are reported in [23], where a data mislabelling problem in real-network traffic datasets is identified. The model for ensuring fairness and transparency in intrusion detection systems is reported in [24]. Although a limited number of papers [23,24], discuss the data-specific issues, in the majority of work reported in the literature does not address the problem of data imbalance. More importantly, the use of Generative AI for generating synthetic data is not explored extensively. 

To that end, in this work, the machine learning is employed for experimentation with real data and synthetic data (created using GANs). GANs consist of two neural networks, a generator and a discriminator, that work against each other and, in the process, generate synthetic data that closely resembles real data [25]. However, there is significant difference the data distributions in both real and synthetic data. This can be considered as bias. In this work, we propose the adversarial debiasing techniques to mitigate and minimize bias in the training process, improving the model's accuracy and adaptability to changing attack patterns. Experiments comparing models trained on real, synthetic, and combined datasets emphasize the important role of debiasing in improving DDoS detection. 

In summary, this study makes the following key contributions:

\begin{enumerate}
    \item A novel, bias-aware detection framework that integrates adversarial debiasing with synthetic data generation using Generative Adversarial Networks (GANs) to significantly improve DDoS detection is introduced. By identifying and neutralizing the influence of sensitive attributes - such as packet direction and size - our approach reduces systemic bias in model predictions. This ensures improved fairness, enhances model generalization, and equips the detection system to adapt to evolving attack strategies in dynamic network environments.
    \item The value of GAN-generated synthetic data in overcoming inherent challenges of real-world datasets, including class imbalance, data sparsity, and the limited representation of rare attack scenarios is clearly demonstrated. The fusion of real and synthetic data in model training substantially boosts robustness and accuracy, while also improving performance on unseen traffic patterns. This work establishes a scalable, adaptable baseline for future research into intelligent, fair, and resilient cybersecurity systems.
    \item This study pioneers the integration of an end-to-end adversarial debiasing pipeline in DDoS detection, combining data-level and model-level interventions to systematically minimize bias from both synthetic and real network traffic sources - a methodological enhancement rarely explored in existing literature.
    \item A phase-based evaluation design that allows for a detailed analysis of model behavior across training, retraining, and inference stages is adopted. This structure reveals critical insights into how synthetic data and debiasing influence generalization, offering practical guidance for deploying adaptive security solutions in real-world settings.

\end{enumerate}

The structure of this paper is as follows: Section 1 provides a brief introduction to the problem statement and relevant information about the related domains. Section 2 presents the literature review of existing work and challenges in DDoS attack detection, machine learning models, and synthetic data generation. Section 3 explains the system design, including the integration of real and synthetic data, the use of adversarial debiasing, and the Random Forest model. Section 4 describes the methodology, outlines the formation of the dataset, the process of synthetic data generation using GANs, the implementation of adversarial debiasing, and the approach followed in the model's training and testing phases. The results and analysis are presented in Section 5, comparing the performance of models trained on real, synthetic, and the mix of the datasets, as well as discussing the effects of different steps of adversarial debiasing on detection accuracy, false positive rates, and adaptability. Section 6 concludes the manuscript with a summary of the findings, limitations, and challenges and offers future research directions. This is followed by the list of statements and declarations, and the list of references.

\section{Literature Review}
This section presents the related work reported in the literature. Sambangi et al [26] considered the problem of identifying DDoS attacks as a classification problem in machine learning (ML). Detecting DDoS attacks presents significant challenges, especially in cloud computing environments, due to the computational complexity involved. A notable challenge in evaluating detection and distinguishing them from Flash Events (FEs) stems from the limited availability of real-world traffic traces suitable for comprehensive assessment. The CICIDS 2017, a popular benchmark dataset, is commonly employed to study and address the challenges of DDoS attack detection in cloud environments [27]. Götte et al. [28] have reported the importance of addressing bias in ML models for ensuring fairness and generalizability [29]. Adversarial debiasing offers a promising approach to mitigate bias by incorporating an adversary that seeks to predict sensitive attributes from the model's output. By training the model to minimize the adversary's performance, the aim is to remove or reduce correlations between the model's predictions and protected attributes. While this can be effective, its impact on performance depends on the characteristics of the data, like measurement accuracy and the representation of relevant subgroups. Hence, carefully evaluating the suitability of the training data and the potential trade-offs between fairness and accuracy is essential when applying adversarial debiasing

Elsayed, et al. [29] have reported the effectiveness of ensemble learning improves the performance and robustness of ML models, particularly neural networks, which are known for their high variance [29]. By combining predictions from multiple models, ensemble methods reduce the prediction variance and enhance the generalization capability, mitigating the risk of overfitting. Methods like bagging, boosting, and model averaging present us with different mechanisms for creating diverse ensembles that contribute to improved predictive accuracy. The choice of ensemble technique depends on certain specific characteristics of the problem and the desired trade-offs between computational cost and performance enhancement.

Kumar and Sinha [30] discussed how the development of systems utilizing ML and data mining techniques offers an approach to addressing the issue of network intrusion detection [30]. However, numerous Intrusion Detection Systems (IDS) grapple with a high frequency of false alarms and fail to identify intrusions. Brownlee [31] wrote about how deep learning, particularly Recurrent Neural Networks (RNNs), holds promise in enhancing the accuracy and effectiveness of anomaly detection systems. RNNs hold the ability to process sequential data, and therefore are well-suited for analyzing network traffic patterns and identifying anomalies that may indicate DDoS attacks [31]. The scarcity of publicly available datasets has led researchers to simulate their data for experimentation, but such datasets miss out on encompassing the intricacies of real-world network behavior. Despite these challenges, the use of DL and RNNs in the form of RNN-autoencoders demonstrates the potential for improving attack detection in SDN or IoT environments by learning features from network traffic data and achieving higher accuracy compared to traditional ML methods [32,33]. 

In [3] a hybrid, incremental-learning framework for DDoS detection using supervised classifiers is discussed. The proposed system splits processing between client and proxy and applies multiple learners (e.g. Naïve Bayes, decision trees, MLP, k-NN) to detect attacks, with Random Forest emerging as the strongest performer. This confirms that ensemble methods like RF are effective in DDoS classification. However, this hybrid approach focuses on accuracy and processing speed in streaming data rather than on data imbalance or bias. The paper does not employ any generative data augmentation or adversarial training; it relies solely on the original (often imbalanced) samples. In contrast, our work explicitly generates synthetic benign/attack samples with GANs to balance the data and applies adversarial debiasing to mitigate feature biases – aspects not addressed by Hosseini and Azizia’s traditional hybrid model. Qing et al. (2024)[5] explicitly tackle the data imbalance problem in IoT-driven DDoS detection. They identify two extreme scenarios – low-rate versus high-rate DDoS attacks – and show that classifiers suffer when cross-evaluated across such varied datasets. To counter this, they experiment with preprocessing techniques and conclude that oversampling should be used for low-rate (rare) attacks and undersampling for high-rate attacks, even suggesting optimized SMOTE oversamplers per model. In essence, they recognize that “two distinct scenarios of data imbalance” exist and that addressing both is crucial. This finding supports our premise that imbalanced attack flows must be mitigated for generalization. However, Qing et al. use conventional resampling (SMOTE variants) rather than generative models, and they do not consider adversarial methods or fairness. In this proposed work, GAN-based synthetic data strategy is employed. This offers a more flexible way to bolster rare classes without discarding real data, and the adversarial debiasing approach proposed here specifically targets bias in packet features. These innovations go beyond the traditional oversampling solutions of Qing et al. and aim to improve both performance and model fairness. In [2], a broad benchmarking of ML classifiers for DDoS detection is conducted. Using four diverse public datasets (e.g. CICDDoS19, backscatter TCP/ICMP, SDN logs) they compare algorithms like CatBoost, logistic regression, gradient boosting and Naïve Bayes on accuracy, precision, recall and F1 score. Their comparative study identifies which classifiers perform best overall, providing a useful overview of detection accuracy across data sources, however, they do not address data bias or imbalance at all. They focus on raw performance metrics and do not include Random Forests in their comparison (despite ensemble methods being strong in literature). Importantly, no synthetic data augmentation or adversarial robustness techniques are used. Our work complements this by not just evaluating a model’s accuracy, but actively improving generalization through GAN-augmented data and reducing bias via adversarial training. Thus, whereas Agrawal et al. report on which off-the-shelf algorithms do well, we introduce methodological novelties to overcome challenges (imbalance and bias) that their evaluation assumes. Novaes et al. (2021)[6] propose an adversarial deep-learning system for DDoS detection in software-defined networks. They employ a Generative Adversarial Network (GAN) framework to not only detect DDoS flows but also to simulate adversarial DDoS traffic for training. Through adversarial training, their model becomes “less sensitive to adversarial attacks” on the network, improving robustness against evasion. This work is closely related to our use of GANs, as it shows that generative models can enhance NIDS performance. The difference is in the goal: Novaes et al. use adversarial examples to harden a deep model, whereas we use adversarial networks to debias a Random Forest. Furthermore, Novaes et al. do not address class imbalance explicitly and focus on robustness to malicious perturbations rather than fairness. Our method, by contrast, integrates GAN-generated benign/attack samples to rebalance training data, and uses adversarial debiasing to remove unwanted correlations in features, which is an aspect absent in the SDN-focused, adversarial-robust design of Novaes et al. In [7] a conceptual survey reframing adversarial ML as a tool for “social good” is presented. They introduce the notion of AdvML4G (Adversarial ML for Social Good), where the “adversary” acts as an ally to produce fairer, safer systems. Their taxonomy explicitly highlights fairness, transparency and trustworthiness as desirable outcomes of repurposed adversarial attacks. This perspective provides valuable motivation for our adversarial debiasing approach: it aligns with the idea that adversarial training can be used to mitigate biases rather than just to attack models. However, Al-Maliki et al. remain high-level and do not tackle any specific domain; they offer no concrete algorithms for DDoS detection or handling imbalanced security data. In other words, while they champion the philosophy of using adversarial methods to improve fairness, our work implements this philosophy in the concrete setting of network security. We translate their broad vision into practical debiasing of a DDoS classifier, whereas their paper primarily justifies why such an approach would be worthwhile. Huang et al. (2024)[8] provide a broad framework on security considerations for generative AI. Their compendium covers threats to generative models (inversion, adversarial, backdoors) and also includes sections on “Model Debiasing and Fairness”. This underscores that even in generative AI, ensuring fairness is a recognized concern. While this work is not about DDoS, it situates our research within the larger context that generative models and adversarial techniques are increasingly studied for security tasks. The book’s coverage of debiasing indicates that mitigating algorithmic bias is timely, but it lacks concrete methods for DDoS or network traffic data. In particular, Huang et al. do not discuss using GANs to synthesize traffic or employing adversarial networks to remove packet-feature bias. Thus, our research can be seen as a domain-specific realization of themes raised by Huang et al.: we apply the book’s general insights about generative models and fairness to the practical problem of DDoS detection, filling the gap between theory and applied cyber security practice. [4] explore DDoS detection with a simple feedforward deep neural network trained on packet-level data. They show that such a DNN can learn to classify flows as benign or attack, achieving high accuracy on their chosen dataset. This confirms that even standard deep models can be effective for DDoS recognition. However, their setup uses relatively “small samples” of data and does not attempt any data augmentation or balancing; the issue of class imbalance is not addressed. They also do not consider adversarial robustness or fairness at all. Compared to this work, [4] represent a baseline deep-learning approach: it lacks the synthetic-data generation and adversarial debiasing that we use to improve generalization. In summary, Cil et al.[4] demonstrate the feasibility of machine learning for DDoS detection, but they do not incorporate any of the bias- or imbalance-mitigation strategies that are novel in our method.

Table~\ref{tab:table1} shows the summary of the literature review with a specific focus on the approach, the datasets used, and the results obtained.

\begin{table}[H]
\centering
\caption{Literature Review Summary}
\label{tab:table1}
\begin{tabularx}{\textwidth}{| X | X | X | X |}
\hline
\textbf{Approach} & \textbf{Dataset} & \textbf{Performance} & \textbf{Key Findings and Limitations} \\
\hline
Feed-forward deep neural network [4] & CICDDoS-2019 & Accuracy improvement & (+) Higher accuracy \newline (−) Small imbalanced dataset \\
\hline
Multiple Linear Regression [26] & CIC-IDS 2017 & 73.79\% & (+) Feature selection via info gain \newline (−) Single day analysis \\
\hline
MLP + Debiasing [27] & Simulated data & 0.005 mean improvement & (+) Tested multiple scenarios \newline (−) Convergence issues \\
\hline
RNN + AutoEncoder [28] & CICDDoS-2019 & 99\% & (+) High accuracy \newline (−) Single dataset evaluation \\
\hline
WCGAN + XGBoost [29] & NSL-KDD, UNSW-NB15, BoT-IoT & Outperformed DGM baseline & (+) Tackles data imbalance \newline (−) Risk of GAN overfitting \\
\hline
GP-WGAN [30] & NSL-KDD, CICIDS2017 & MSE: 0.10, AUC: 75\% & (+) Generates synthetic attacks \newline (−) Limited sample size (2000) \\
\hline
Adversarial Debiasing [33] & Not available & Not available & (+) Improves model reliability \newline (−) Limited to static analysis \\
\hline
Various ML approaches (ANN, RF, etc.) [35] & CIC-IDS 2017 & ANN had the best performance & (+) Comprehensive method comparison \newline (−) Limited dataset scope \\
\hline
EMCSVM & Custom \newline (14 attributes) & Outperformed the standard SVM & (+) Real-time detection \newline (−) High computation needs \\
\hline
\end{tabularx}
\end{table}
\section{System Design}

The proposed framework is divided into 4 phases, as illustrated in Fig. 1 and Fig. 2 

\subsubsection{Phase 1: Training the Initial Classification Model} 
The process began by training the baseline classification model using the original real-world dataset. This data set contained distributed denial of Service (DDoS) and benign network packets and served as a reference for subsequent stages.

\subsubsection{Phase 2: Generating Synthetic Data with GANs} 
A GAN model was trained and used to generate synthetic data samples. The GAN was trained on the real dataset to produce synthetic packets, which were divided into three subsets. Subsequently, 75,000 DDoS and 75,000 BENIGN packets were generated, merged, and saved into a single CSV file, which had a total of 150,000 synthetic samples.

\subsubsection{Phase 3: Retraining the Classification Model with Combined Datasets} 
The original classification model was retrained on a combination of the synthetic dataset, which had 150,000 samples, and an equal number of samples from the original dataset. This created a balanced dataset with a ratio of 1:1 of DDoS to BENIGN packets, resulting in a total of 300,000 samples for model training. The goal of this step was to integrate synthetic data to minimize bias and enhance the model's ability to generalize effectively.

\subsubsection{Phase 4: Inference and Evaluation of the Debiased Model} 
A separate holdout dataset, consisting of 2,000 samples generated in Phase 2, was used to assess the performance of the retrained model. The model achieved an accuracy of 72.60\%, indicating that incorporating synthetic data and debiasing has improved its predictive effectiveness. In this study, each phase of data preprocessing, GAN-based synthetic data generation, model training, and inference was developed as an independent module to ensure flexibility. Distributed computing frameworks such as TensorFlow were used to train the model on a dataset of 300,000 samples, which efficiently optimized the process.

\begin{figure}[!h]
    \centering
    \includegraphics[width=1\linewidth]{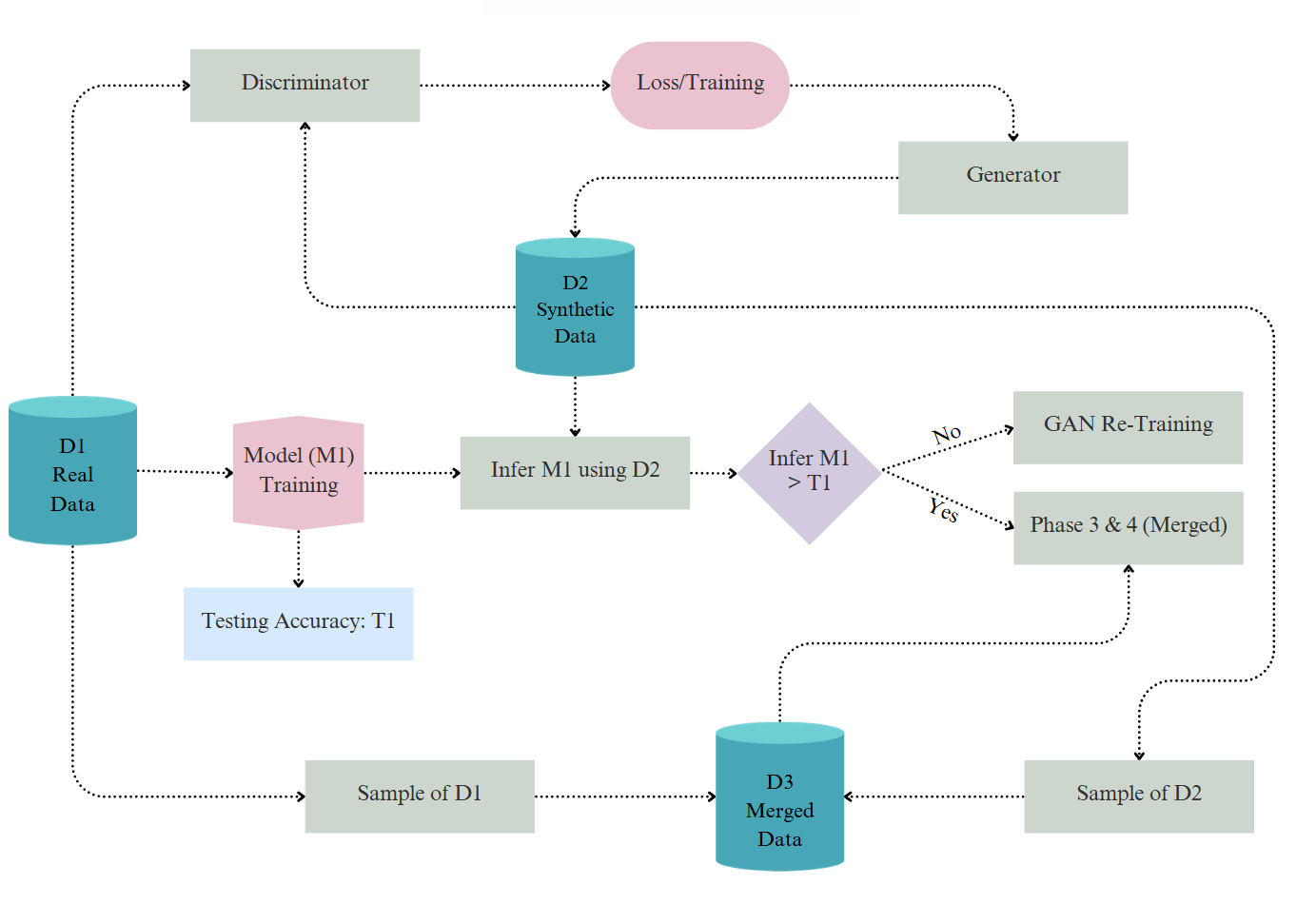}
    \caption{Project system design for Phase 1 and 2}
    \label{fig:FigureOne}
\end{figure}
\begin{figure}[!h]
    \centering
    \includegraphics[width=1\linewidth]{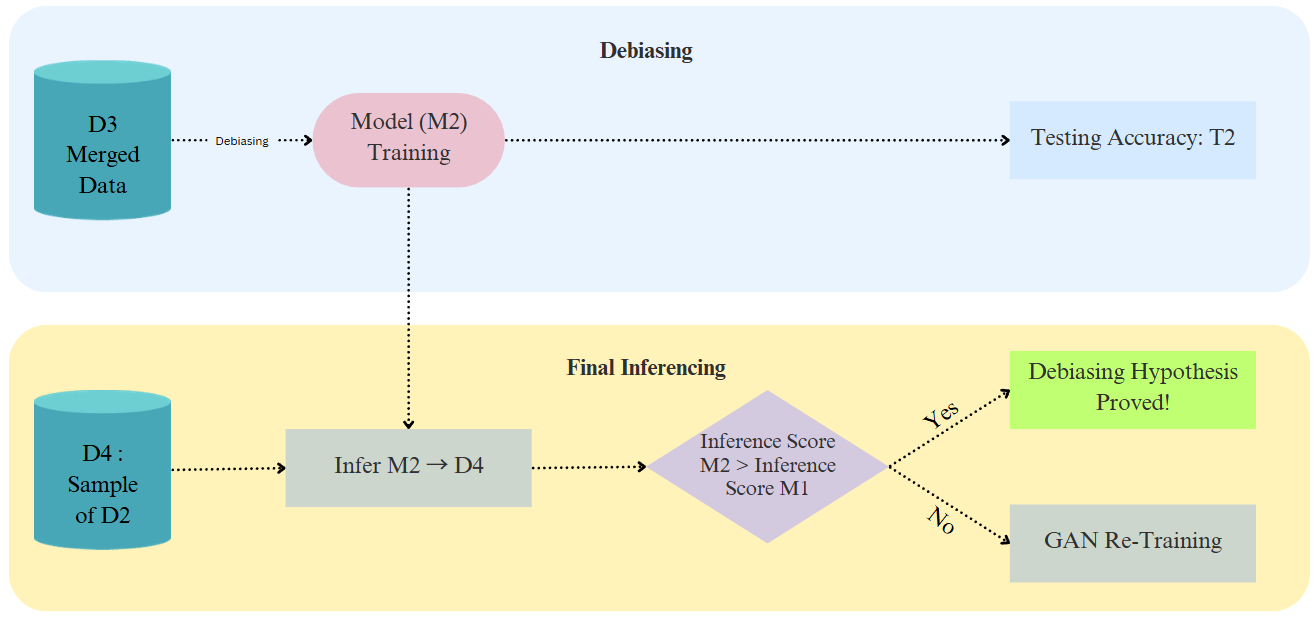}
    \caption{Project system design for Phase 3 and 4}
    \label{fig:FigureTwo}
\end{figure}

\section{Methodology}
This section outlines the methodology followed in data preprocessing, model training, and evaluation to rigorously assess the effectiveness of our proposed framework. It describes how real-world and GAN-generated synthetic data were integrated to improve detection capabilities and how adversarial debiasing was implemented to reduce systemic bias. The methodology is structured into several key phases, beginning with data analysis and culminating in performance evaluation, providing a comprehensive approach to building a resilient DDoS detection model.
\subsection{Data Analysis}
This study integrates real and synthetic data to improve the accuracy and robustness of a DDoS detection model. The real dataset used is the CIC-IDS2017 dataset [27]. This dataset includes labeled instances of both normal traffic and DDoS attacks. The dataset used for training the GAN had 85 features and 225,745 rows. Important features were selected by analyzing the network-related columns to prepare the data. Since DDoS attacks affect network traffic, features like high forward (FWD) and backward (BWD) packet counts were identified as key indicators of unusual network behavior. After calculating the FWD and BWD packet counts on a cumulative sum graph against the flow of the packets, as seen in Fig. 3.

It has been observed that the spikes in the graphs of the blue line indicate the sudden increment in the number of packet counts. Ideally, over a TCP handshake, the FWD packets should be almost equal to the BWD packet counts. The columns that had an effect on the network were as follows: 

\textit{Flow Duration, Total Fwd Packets, Total Backward Packets, Total Length of Fwd Packets, Total Length of Bwd Packets, Fwd Packet Length Max, Fwd Packet Length Min, Fwd Packet Length Std, Bwd Packet Length Min, Flow Bytes/s, Flow Packets/s, Flow IAT Mean, Flow IAT Std, Flow IAT Max, Flow IAT Min, Fwd IAT Total, Fwd IAT Max, Fwd IAT Min, Bwd IAT Total, Bwd IAT Max, Bwd IAT Min, Min Packet Length, Max Packet Length, Packet Length Mean, Packet Length Std, Packet Length Variance, FIN Flag Count, SYN Flag Count, RST Flag Count, PSH Flag Count, ACK Flag Count, URG Flag Count, CWE Flag Count, ECE Flag Count, Down/Up Ratio, Average Packet Size, Avg Fwd Segment Size, Avg Bwd Segment Size, Fwd Header Length.1, Subflow Fwd Bytes, Subflow Bwd Bytes, Active Mean, Active Std, Active Max, Active Min, Idle Mean, Idle Std, Idle Max, Idle Min, Label.}

The above parameters (features) were used to train the classification model and the GAN. The cleaned dataset had 225745 rows and 50 columns, ready for model training.

\subsection{Classification}
A key aspect of this study involves accurately predicting whether network packets are instances of a DDoS attack or BENIGN packets. ML models were employed to accomplish this. StandardScaler was used to preprocess the data before inputting it into a machine-learning model that ensures that all features are on an equal scale, preventing any single feature from overshadowing the learning process because of its larger magnitude. 

Here, the performance of five distinct machine-learning algorithms, namely Logistic Regression, Decision Tree, Support Vector Machine (SVM), Random Forest, and K-Nearest Neighbors (KNN), which are predictive models engineered to handle network traffic classification, were evaluated using real baseline data. This analysis aims to provide insights into the relative efficacy of these algorithms in the context of the specified dataset. 

After testing all of the above models, the Random Forest Classifier achieved the highest accuracy of 99.967\% in classifying DDoS and BENIGN network packets. This is because it constructs an ensemble of multiple Decision Trees during the training phase and uses voting for cooperative decision-making, leading to precise outcomes [37].

\subsection{Generative Adversarial Network}
To generate synthetic packets that resemble the DDoS or BENIGN packets that the original dataset had, generative adversarial networks (GANs) were used. They are composed of a generator and a discriminator. The goal is to estimate the potential distribution of real data samples and generate new samples from that distribution [38].

The generator model is trained on the cleaned data's dimensions and consists of Dense layers with Leaky ReLU activations to prevent dead neurons [39]. Batch Normalization stabilizes training by normalizing activations. The model starts with 128 neurons, increasing to 512 to capture complex patterns, and ends with a tanh activation to map outputs to [-1, 1], aligning with real data scaling. This design enables the generator to produce realistic synthetic samples [40].

The discriminator model uses the same input dimensions and processes data through Dense layers that reduce feature representation. It starts with 512 neurons for complex pattern detection, followed by 256 neurons for refinement. Leaky ReLU is applied after each layer for maintaining gradients and addressing inactive neurons. This architecture ensures the discriminator effectively distinguishes real from generated data, pushing the generator to improve over time.

The discriminator is compiled with a binary cross-entropy loss and Adam optimizer, stabilizing training with low learning rates and momentum settings [41]. During each epoch, the generator creates synthetic data using random noise as input, and the discriminator is trained on both real and synthetic data, reinforcing its ability to identify genuine samples. Simultaneously, the generator learns from the feedback to improve its outputs to fool the discriminator. Training losses and accuracy are tracked, and checkpoints are used to save the models at intervals. After training, the generator produces synthetic DDoS data, which is combined with BENIGN data and converted into a final dataset that is saved for analysis.

\subsection{Inferencing}
The Random Forest model, trained on real-world network traffic data, served as a baseline to evaluate the ability of a model trained on authentic data to classify synthetic samples (DDoS or BENIGN) generated by the GAN. This inference aimed to assess the quality of GAN-generated data and detect potential biases in the model towards real data.

In contrast to poor performance which suggests shortcomings in the GANs capacity to replicate the complexity and diversity of real data, high performance on synthetic data shows that the GAN successfully captured the underlying patterns of real-world traffic. The model’s accuracy on synthetic data was only 50\% which was much worse than its performance on real data. This brought to light the difficulties in producing high-quality synthetic data that accurately captures the complexities of actual network traffic. 

The model performed poorly on synthetic data for several reasons. Even though the GAN was created to produce realistic-looking synthetic network packets the predictive accuracy of the model may be impacted by minute variations in the data distribution between the real and synthetic datasets. Furthermore, synthetic data is frequently less noisy and cleaner than real-world data. Because it might not be strong enough to cope with the lack of noise in the synthetic dataset a model trained on real data might have trouble with this discrepancy. 

The model's intrinsic bias towards the real-world data it was trained on however is the main cause of this subpar performance. This bias results from the unique properties of real data which the GAN might not be able to fully replicate. As a result, the model's capacity to generalize to unknown synthetic data is restricted and its accuracy is consequently low. 

De-biasing the model was necessary to overcome these obstacles and make it function well on both synthetic and real network traffic. To develop a more robust and dependable model for identifying DDoS attacks across various data sources debiasing techniques were used. 

\begin{figure}[!h]
    \centering
    \includegraphics[width=1\linewidth]{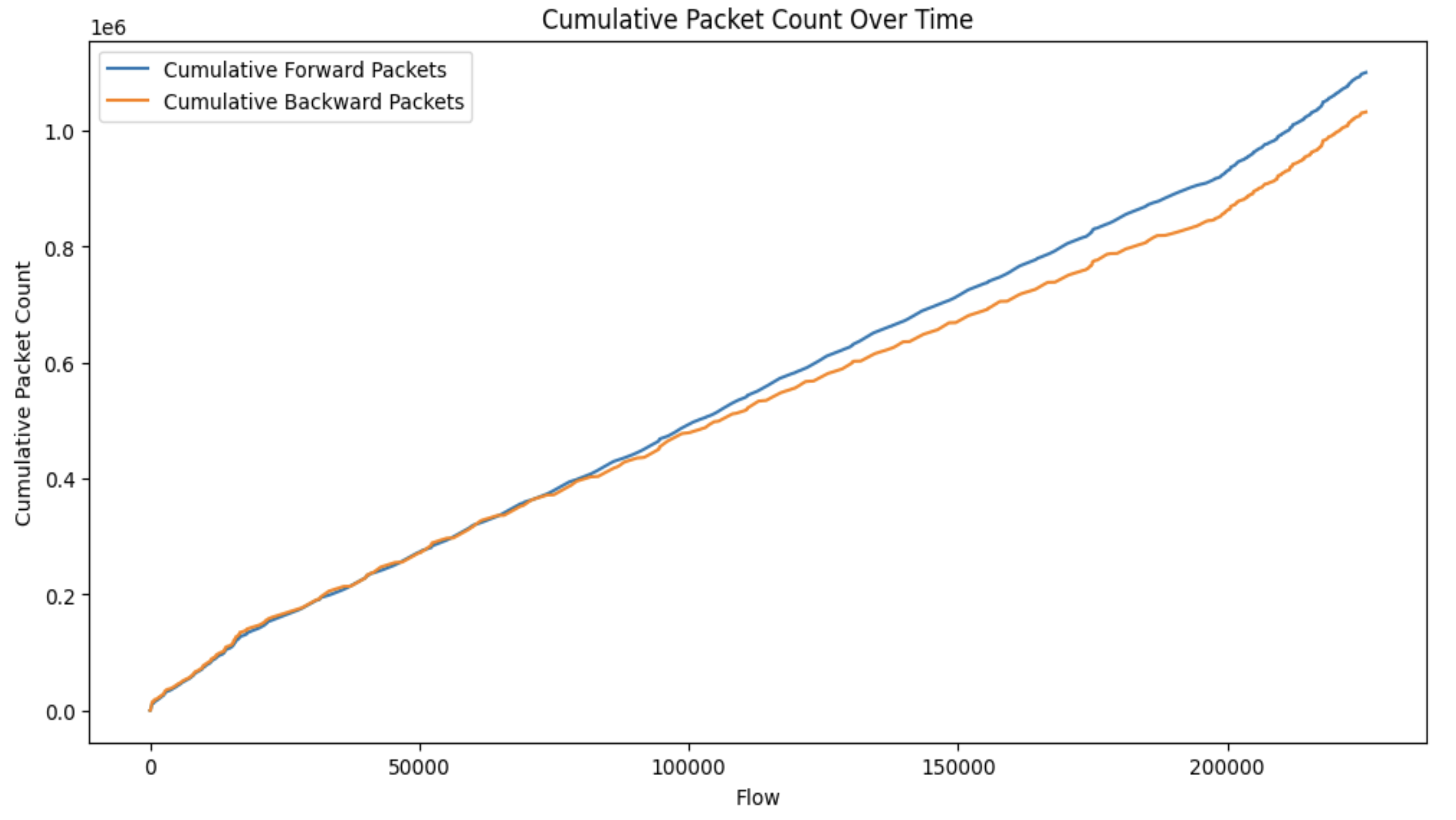}
    \caption{Cumulative forward and backward packet count}
    \label{fig:Cummulative}
\end{figure}

\subsection{Debiasing}
Biases resulting from imbalances or insufficiencies in the training data can affect machine learning models causing skewed predictions and affecting the model's generalizability [42]. Inaccurate attack patterns and unequal volumes of malicious and legitimate traffic are examples of biases in DDoS attack detection that can make it difficult to detect attacks. During the GAN synthetic data generation process adversarial debiasing was used to address these problems. By lowering bias in the synthetic dataset this method gave attacks a more balanced portrayal. The Random Forest model was then enhanced and improved by integrating the debiased synthetic data produced by the GAN into the training procedure. This method reduced biases and enabled the model to detect new attack patterns, making the proposed DDoS detection model more robust and less sensitive to biases.

\section{Result and Analysis}

An ensemble of five models was trained and tested with the results observed in Table 2.

The Random Forest classifier outperformed all the others in classifying between BENIGN and DDoS network packets, achieving an accuracy of 99.967\%. The confusion matrix (Fig. 3) visualizes the robustness of the model. In the subsequent phase of the study, the generated synthetic network data had a cosine similarity score of 0.8031, which is similar to real network traffic data. However, when the synthetic data was inference with the classification model, a poor accuracy of 50.00\% proved that inherent bias exists. 

The model was then retrained on an augmented dataset combining both real and synthetic data, resulting in an increased testing accuracy of 99.987\%. The confusion matrix for the retrained Random Forest model (Fig. 5) shows a decrease in false negatives and false positives compared to the initial baseline model (Fig. 4), highlighting enhanced robustness by the model in the results. The retrained model, with its inference done using synthetic packets, achieved an increased accuracy of 72.60\%, surpassing the previous metric where the model was only trained on real data.

This highlights the effectiveness of debiasing in improving a model’s ability to handle unseen data. Fig. 6 demonstrates graphically throughout the phases how the models have improved in their accuracy. Phase 1 and Phase 2 highlights the baseline model where the model had an accuracy of 99.967\% when tested on real baseline data, while it performed poorly to just 50\% accuracy when inferenced with synthetic unseen data. Conversely, when retraining the model with a mix of real and synthetic data, the testing accuracy improved to 99.987\% in Phase 3 and inferencing it with unseen synthetic data in Phase 4 indicated an observable increase to 72.6\% in accuracy.

\begin{table}
\centering
\caption{A table representing the testing accuracy achieved by different ML models in classifying between DDoS and benign network packets}
\label{tab:metrics}
\begin{tabular}{| l | l | l |}
\hline
No.

 & Machine Learning Model & Testing Accuracy (\%) \\
\hline
1. & Logistic Regression & 99.84 \\
\hline
2. & Decision Tree & 99.96 \\
\hline
3. & Support Vector Machine & 99.84 \\
\hline
4. & Random Forest & 99.96 \\
\hline
5. & K-Nearest Neighbors & 99.93 \\
\hline

\end{tabular}

\end{table}

\section{Conclusion and Future Work}
The proposed framework combining synthetic data generation using GANs with adversarial debiasing and ensemble learning has shown significant promise in enhancing the detection of DDoS attacks. By examining the performance across various phases of model training and evaluation, this study provides insights into the role of fairness-aware learning and data augmentation in cyber security. The subsections that follow outline the major contributions of this study, acknowledge its current limitations, and propose actionable directions for future research and development.
\subsection{Main Contributions and Novelty of Study}
\begin{enumerate}
    \item This study introduces a novel framework that combines GAN-generated synthetic data with adversarial debiasing techniques to improve the detection of DDoS attacks.
    \item It demonstrates that retraining a Random Forest classifier with a mixture of synthetic and real-world data significantly enhances detection accuracy and robustness.
    \item The approach effectively mitigates model bias by focusing on sensitive attributes such as packet counts and byte lengths, leading to fairer and more generalized predictions.
    \item The proposed system achieved near-perfect accuracy on benchmark data and a notable performance gain (+22.60\%) on unseen synthetic datasets.
    \item This research validates the use of GANs for realistic traffic data simulation and emphasizes adversarial debiasing as a powerful tool for cyber security applications.
\end{enumerate}
\begin{figure}[!h]
    \centering
    \includegraphics[width=1\linewidth]{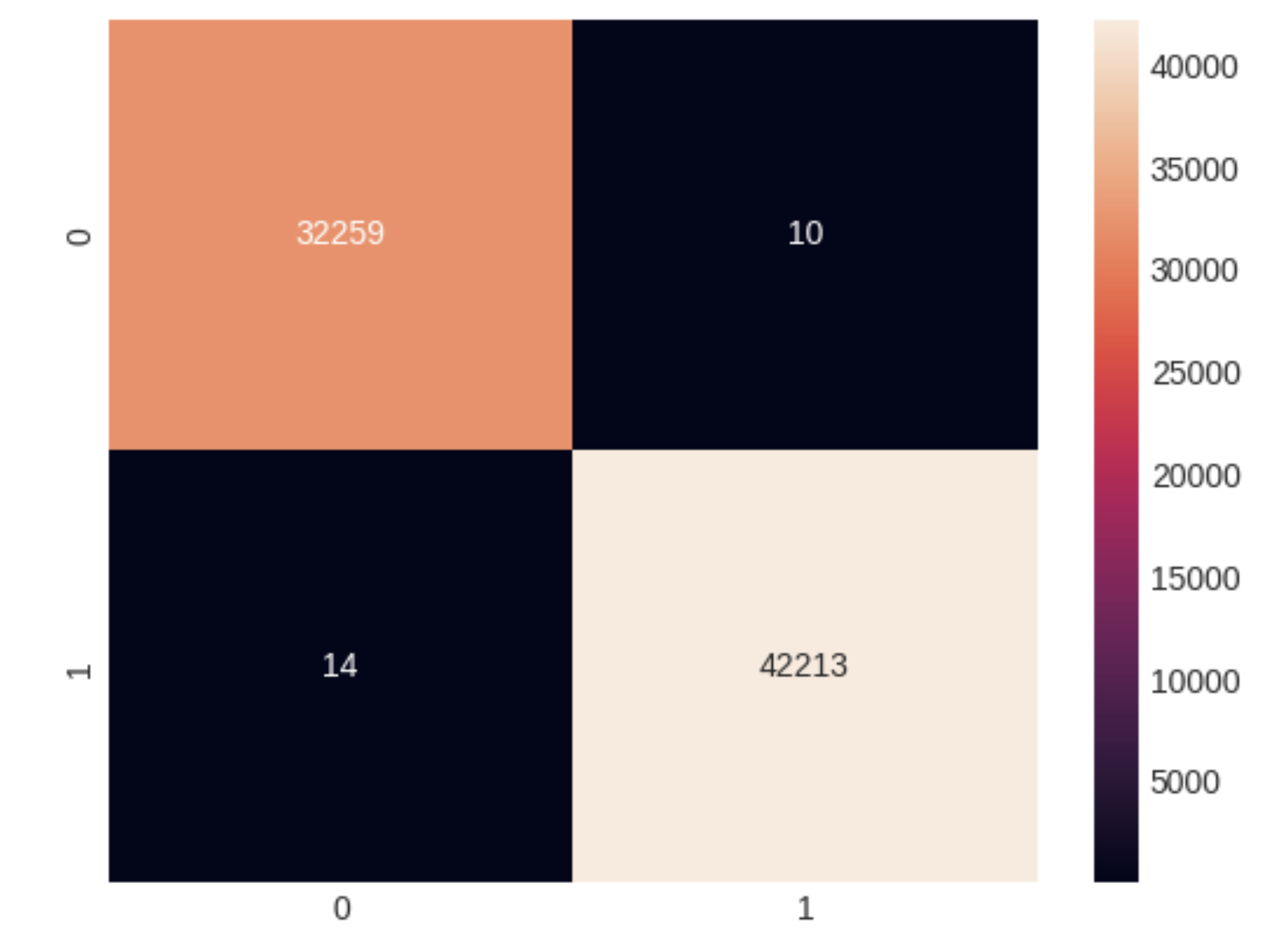}
    \caption{Confusion matrix for the baseline Random Forest model}
    \label{fig:FigureFour}
\end{figure}
\begin{figure}[!h]
    \centering
    \includegraphics[width=1\linewidth]{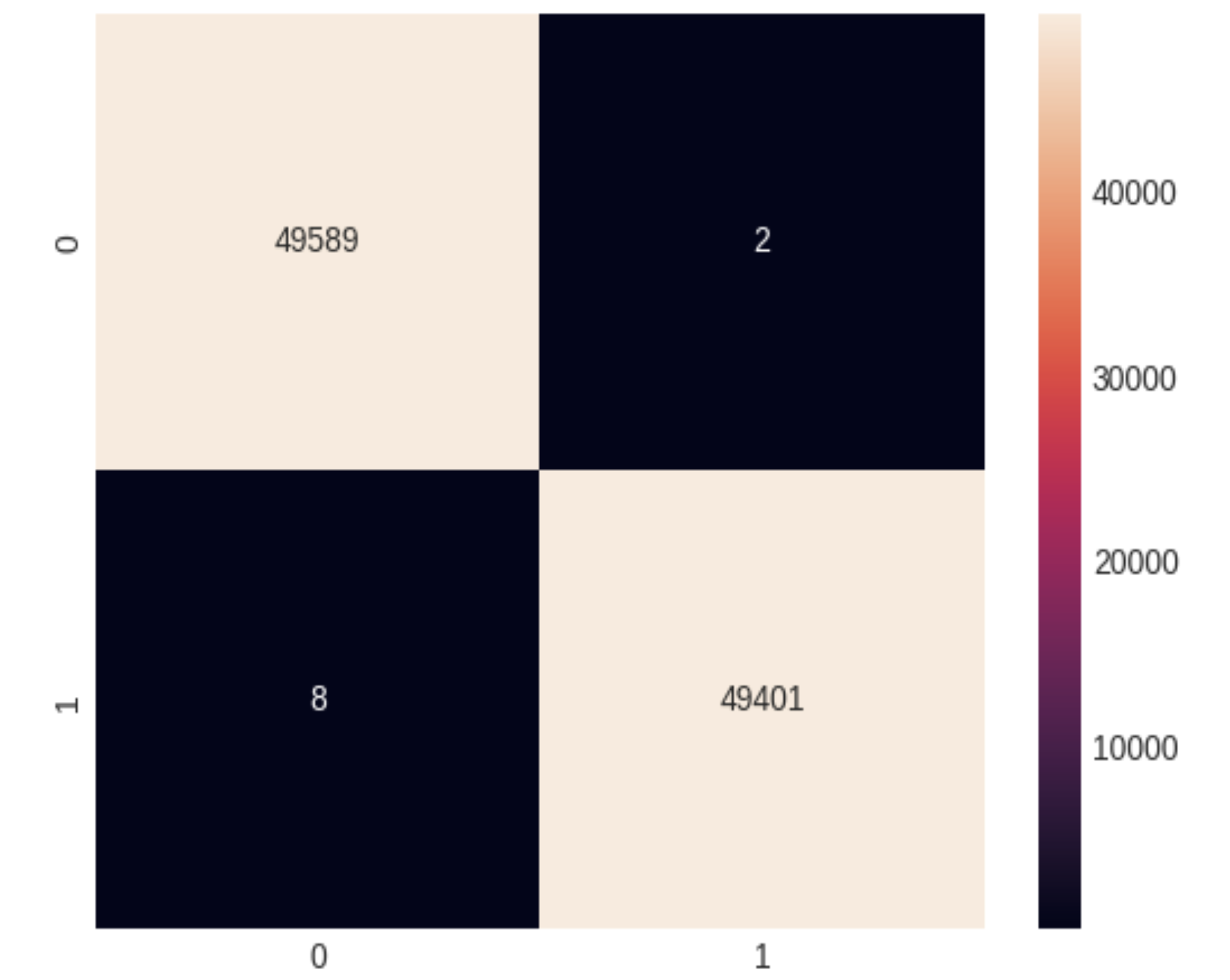}
    \caption{Confusion matrix for the baseline Random Forest model}
    \label{fig:FigureFive}
\end{figure}
\subsection{Limitation of Study}
\begin{enumerate}
    \item The model was primarily tested on a controlled synthetic dataset, which may not fully represent real-time network complexity.
    \item The GAN-generated data, while similar, still exhibited distributional gaps compared to real-world traffic, affecting baseline performance before retraining. 
\item The study focused only on Random Forest and did not compare performance against advanced deep learning models such as CNNs or RNNs. 
\item Adversarial debiasing was applied to a limited set of features, potentially overlooking biases in other parts of the data. 
\item Real-time deployment and evaluation were not conducted, limiting insight into how the model would perform under live network conditions.
\end{enumerate}

\subsection{Recommendations for Future Research}
\begin{enumerate}
    \item Future work should include experimentation with advanced deep learning architectures like RNNs and CNNs to capture spatial and temporal dynamics in traffic data. 
    \item Expanding adversarial debiasing to a broader set of features may further improve model fairness and robustness. 
\item Evaluation should be extended to real-time environments with live data streams to test adaptability under real-world attack conditions. 
\item Applying the framework to other types of cyber threats, such as ransomware or phishing, can assess its generalizability across the cybersecurity spectrum. 
\item Incorporating federated learning or edge-computing capabilities could help build scalable and privacy-preserving DDoS detection systems.
\end{enumerate}

\section{Acknowledgment}
This article does not contain any studies with human participants or animals performed by any of the authors. This manuscript is the authors’ original work, and the content of this paper has not been copied from elsewhere. All authors have checked the manuscript and have agreed to this submission. Availability of data and materials: This study uses CICDDoS-2017 dataset which is publicly available and can be accessed at https://www.unb.ca/cic/datasets/ids-2017.html. The second dataset used is the synthetic dataset developed as part of this work that will be available upon request. During the preparation of this work the author(s) used ChatGPT in order to improve grammar and sentence structure. After using this tool/service, the author(s) reviewed and edited the content as needed and take(s) full responsibility for the content of the published article.

\end{document}